

\font\twelverm=cmr10 scaled 1200    \font\twelvei=cmmi10 scaled 1200
\font\twelvesy=cmsy10 scaled 1200   \font\twelveex=cmex10 scaled 1200
\font\twelvebf=cmbx10 scaled 1200   \font\twelvesl=cmsl10 scaled 1200
\font\twelvett=cmtt10 scaled 1200   \font\twelveit=cmti10 scaled 1200
\font\twelvesc=cmcsc10 scaled 1200  
\skewchar\twelvei='177   \skewchar\twelvesy='60
\def\twelvepoint{\normalbaselineskip=12.4pt plus 0.1pt minus 0.1pt
  \abovedisplayskip 12.4pt plus 3pt minus 9pt
  \belowdisplayskip 12.4pt plus 3pt minus 9pt
  \abovedisplayshortskip 0pt plus 3pt
  \belowdisplayshortskip 7.2pt plus 3pt minus 4pt
  \smallskipamount=3.6pt plus1.2pt minus1.2pt
  \medskipamount=7.2pt plus2.4pt minus2.4pt
  \bigskipamount=14.4pt plus4.8pt minus4.8pt
  \def\rm{\fam0\twelverm}          \def\it{\fam\itfam\twelveit}%
  \def\sl{\fam\slfam\twelvesl}     \def\bf{\fam\bffam\twelvebf}%
  \def\mit{\fam 1}                 \def\cal{\fam 2}%
  \def\sc{\twelvesc}               \def\tt{\twelvett}
  \def\sf{\twelvesf}
  \textfont0=\twelverm   \scriptfont0=\tenrm   \scriptscriptfont0=\sevenrm
  \textfont1=\twelvei    \scriptfont1=\teni    \scriptscriptfont1=\seveni
  \textfont2=\twelvesy   \scriptfont2=\tensy   \scriptscriptfont2=\sevensy
  \textfont3=\twelveex   \scriptfont3=\twelveex  \scriptscriptfont3=\twelveex
  \textfont\itfam=\twelveit
  \textfont\slfam=\twelvesl
  \textfont\bffam=\twelvebf \scriptfont\bffam=\tenbf
  \scriptscriptfont\bffam=\sevenbf
  \normalbaselines\rm}

\def\beginlinemode{\endmode
  \begingroup\parskip=0pt \obeylines\def\\{\par}\def\endmode{\par\endgroup}}
\def\beginparmode{\endmode
  \begingroup \def\endmode{\par\endgroup}}
\let\endmode=\par
{\obeylines\gdef\
{}}
\def\singlespace{\baselineskip=\normalbaselineskip}

\def\oneandahalfspace{\baselineskip=\normalbaselineskip
  \multiply\baselineskip by 3 \divide\baselineskip by 2}
\def\doublespace{\baselineskip=\normalbaselineskip \multiply\baselineskip by 2}

\newcount\firstpageno
\firstpageno=2
\footline={\ifnum\pageno<\firstpageno{\hfil}\else{\hfil\twelverm\folio\hfil}\fi}
\def\toppageno{\global\footline={\hfil}\global\headline
  ={\ifnum\pageno<\firstpageno{\hfil}\else{\hfil\twelverm\folio\hfil}\fi}}
\let\rawfootnote=\footnote              
\def\footnote#1#2{{\rm\singlespace\parindent=0pt\parskip=0pt
  \rawfootnote{#1}{#2\hfill\vrule height 0pt depth 6pt width 0pt}}}
\def\raggedcenter{\leftskip=4em plus 12em \rightskip=\leftskip
  \parindent=0pt \parfillskip=0pt \spaceskip=.3333em \xspaceskip=.5em
  \pretolerance=9999 \tolerance=9999
  \hyphenpenalty=9999 \exhyphenpenalty=9999 }
\def\dateline{\rightline{\ifcase\month\or
  January\or February\or March\or April\or May\or June\or
  July\or August\or September\or October\or November\or December\fi
  \space\number\year}}
\def\received{\vskip 3pt plus 0.2fill
 \centerline{\sl (Received\space\ifcase\month\or
  January\or February\or March\or April\or May\or June\or
  July\or August\or September\or October\or November\or December\fi
  \qquad, \number\year)}}

\hsize=6.5truein
\vsize=8.9truein
\parskip=\medskipamount
\def\\{\cr}
\twelvepoint            
\doublespace            
\overfullrule=0pt       



\def\title                      
  {\null\vskip 3pt plus 0.2fill
   \beginlinemode \doublespace \raggedcenter \bf}

\def\author                     
  {\vskip 3pt plus 0.2fill \beginlinemode
   \singlespace \raggedcenter\sc}

\def\affil                      
  {\vskip 3pt plus 0.1fill \beginlinemode
   \oneandahalfspace \raggedcenter \sl}

\def\abstract                   
  {\vskip 3pt plus 0.3fill \beginparmode
   \singlespace ABSTRACT: }

\def\endtopmatter               
  {\endpage                     
   \body}

\def\body                       
  {\beginparmode}               

\def\head#1{                    
  \goodbreak\vskip 0.5truein    
  {\immediate\write16{#1}
   \raggedcenter \uppercase{#1}\par}
   \nobreak\vskip 0.25truein\nobreak}

\def\subhead#1{                 
  \vskip 0.25truein             
  {\raggedcenter {#1} \par}
   \nobreak\vskip 0.25truein\nobreak}

\def\beginitems{
\par\medskip\bgroup\def\i##1 {\item{##1}}\def\ii##1 {\itemitem{##1}}
\leftskip=36pt\parskip=0pt}
\def\enditems{\par\egroup}

\def\beneathrel#1\under#2{\mathrel{\mathop{#2}\limits_{#1}}}

\def\refto#1{$^{#1}$}           

\def\references                 
  {\head{References}            
   \beginparmode
   \frenchspacing \parindent=0pt \leftskip=1truecm
   \parskip=8pt plus 3pt \everypar{\hangindent=\parindent}}

\gdef\refis#1{\item{#1.\ }}                     

\gdef\journal#1, #2, #3, 1#4#5#6{               
    {\sl #1~}{\bf #2}, #3 (1#4#5#6)}            

\gdef\refa#1, #2, #3, #4, 1#5#6#7.{\noindent#1, #2 {\bf #3}, #4 (1#5#6#7).\rm}
\gdef\refb#1, #2, #3, #4, 1#5#6#7.{\noindent#1 (1#5#6#7), #2 {\bf #3}, #4.\rm}

\def\endreferences{\body}

\def\figurecaptions             
  {\endpage
   \beginparmode
   \head{Figure Captions}
}

\def\endpage                    
  {\vfill\eject}

\def\endpaper                   
  {\endmode\vfill\supereject}

\def\heading                            
  {\vskip 0.5truein plus 0.1truein      
   \beginparmode \def\\{\par} \parskip=0pt \singlespace \raggedcenter}

\def\subheading                         
  {\vskip 0.25truein plus 0.1truein     
   \beginlinemode \singlespace \parskip=0pt \def\\{\par}\raggedcenter}

\def\tag#1$${\eqno(#1)$$}

\def\align#1$${\eqalign{#1}$$}

\def\aligntag#1$${\gdef\tag##1\\{&(##1)\cr}\eqalignno{#1\\}$$
  \gdef\tag##1$${\eqno(##1)$$}}

\def\overset #1\to#2{{\mathop{#2}\limits^{#1}}}
\def\underset#1\to#2{{\let\next=#1\mathpalette\undersetpalette#2}}
\def\undersetpalette#1#2{\vtop{\baselineskip0pt
\ialign{$\mathsurround=0pt #1\hfil##\hfil$\crcr#2\crcr\next\crcr}}}

\def\ref#1{Ref.~#1}                     
\def\Ref#1{Ref.~#1}                     
\def\[#1]{[\cite{#1}]}
\def\cite#1{{#1}}
\def\(#1){(\call{#1})}
\def\call#1{{#1}}
\def\taghead#1{}
\def\frac#1#2{{#1 \over #2}}
\def\half{{\frac 12}}

\def\12{{1\over2}}

\def\sla{\raise.15ex\hbox{$/$}\kern-.57em}
\def\leaderfill{\leaders\hbox to 1em{\hss.\hss}\hfill}
\def\twiddle{\lower.9ex\rlap{$\kern-.1em\scriptstyle\sim$}}
\def\bigtwiddle{\lower1.ex\rlap{$\sim$}}
\def\gtwid{\mathrel{\raise.3ex\hbox{$>$\kern-.75em\lower1ex\hbox{$\sim$}}}}
\def\ltwid{\mathrel{\raise.3ex\hbox{$<$\kern-.75em\lower1ex\hbox{$\sim$}}}}
\def\square{\kern1pt\vbox{\hrule height 1.2pt\hbox{\vrule width 1.2pt\hskip 3pt
   \vbox{\vskip 6pt}\hskip 3pt\vrule width 0.6pt}\hrule height 0.6pt}\kern1pt}
\def\tdot#1{\mathord{\mathop{#1}\limits^{\kern2pt\ldots}}}

\def\pmb#1{\setbox0=\hbox{#1}%
  \kern-.025em\copy0\kern-\wd0
  \kern  .05em\copy0\kern-\wd0
  \kern-.025em\raise.0433em\box0 }

\catcode`@=11
\newcount\r@fcount \r@fcount=0
\newcount\r@fcurr
\immediate\newwrite\reffile
\newif\ifr@ffile\r@ffilefalse
\def\w@rnwrite#1{\ifr@ffile\immediate\write\reffile{#1}\fi\message{#1}}

\def\writer@f#1>>{}
\def\referencefile{
  \r@ffiletrue\immediate\openout\reffile=\jobname.ref%
  \def\writer@f##1>>{\ifr@ffile\immediate\write\reffile%
    {\noexpand\refis{##1} = \csname r@fnum##1\endcsname = %
     \expandafter\expandafter\expandafter\strip@t\expandafter%
     \meaning\csname r@ftext\csname r@fnum##1\endcsname\endcsname}\fi}%
  \def\strip@t##1>>{}}

\def\citeall#1{\xdef#1##1{#1{\noexpand\cite{##1}}}}
\def\cite#1{\each@rg\citer@nge{#1}}	

\def\each@rg#1#2{{\let\thecsname=#1\expandafter\first@rg#2,\end,}}
\def\first@rg#1,{\thecsname{#1}\apply@rg}	
\def\apply@rg#1,{\ifx\end#1\let\next=\relax
\else,\thecsname{#1}\let\next=\apply@rg\fi\next}

\def\citer@nge#1{\citedor@nge#1-\end-}	
\def\citer@ngeat#1\end-{#1}
\def\citedor@nge#1-#2-{\ifx\end#2\r@featspace#1 
  \else\citel@@p{#1}{#2}\citer@ngeat\fi}	
\def\citel@@p#1#2{\ifnum#1>#2{\errmessage{Reference range #1-#2\space is bad.}%
    \errhelp{If you cite a series of references by the notation M-N, then M and
    N must be integers, and N must be greater than or equal to M.}}\else%
 {\count0=#1\count1=#2\advance\count1 by1\relax\expandafter\r@fcite\the\count0,
  \loop\advance\count0 by1\relax
    \ifnum\count0<\count1,\expandafter\r@fcite\the\count0,%
  \repeat}\fi}

\def\r@featspace#1#2 {\r@fcite#1#2,}	
\def\r@fcite#1,{\ifuncit@d{#1}
    \newr@f{#1}%
    \expandafter\gdef\csname r@ftext\number\r@fcount\endcsname%
                     {\message{Reference #1 to be supplied.}%
                      \writer@f#1>>#1 to be supplied.\par}%
 \fi%
 \csname r@fnum#1\endcsname}
\def\ifuncit@d#1{\expandafter\ifx\csname r@fnum#1\endcsname\relax}%
\def\newr@f#1{\global\advance\r@fcount by1%
    \expandafter\xdef\csname r@fnum#1\endcsname{\number\r@fcount}}

\let\r@fis=\refis			
\def\refis#1#2#3\par{\ifuncit@d{#1}
   \newr@f{#1}%
   \w@rnwrite{Reference #1=\number\r@fcount\space is not cited up to now.}\fi%
  \expandafter\gdef\csname r@ftext\csname r@fnum#1\endcsname\endcsname%
  {\writer@f#1>>#2#3\par}}

\def\ignoreuncited{
   \def\refis##1##2##3\par{\ifuncit@d{##1}%
    \else\expandafter\gdef\csname r@ftext\csname r@fnum##1\endcsname\endcsname%
     {\writer@f##1>>##2##3\par}\fi}}

\def\r@ferr{\endreferences\errmessage{I was expecting to see
\noexpand\endreferences before now;  I have inserted it here.}}
\let\r@ferences=\references
\def\references{\r@ferences\def\endmode{\r@ferr\par\endgroup}}

\let\endr@ferences=\endreferences
\def\endreferences{\r@fcurr=0
  {\loop\ifnum\r@fcurr<\r@fcount
    \advance\r@fcurr by 1\relax\expandafter\r@fis\expandafter{\number\r@fcurr}%
    \csname r@ftext\number\r@fcurr\endcsname%
  \repeat}\gdef\r@ferr{}\endr@ferences}


\let\r@fend=\endpaper\gdef\endpaper{\ifr@ffile
\immediate\write16{Cross References written on []\jobname.REF.}\fi\r@fend}

\catcode`@=12

\citeall\refto		
\citeall\ref		%
\citeall\Ref		%


\def\D{\Delta}
\def\la{\langle}
\def\ra{\rangle}
\def\ria{\rightarrow}
\def\x{{\bar x}}
\def\k{{\bar k}}

\def\a{\alpha}
\def\b{\beta}
\def\U{\Upsilon}
\def\G{\Gamma}
\def\Tr{{\rm Tr}}
\def\ih{{ {i \over \hbar} }}
\def\au{{\underline{\alpha}}}
\def\s{{\sigma}}

\def\pp{{\prime\prime}}
\def\jjh{j\_halliwell@vax1.physics.imperial.ac.uk}


\centerline{\bf Quantum-Mechanical Histories and the
Uncertainty Principle:}
\centerline{\bf I. Information-Theoretic Inequalities.}

\vskip 0.5in
\author J.J.Halliwell\footnote{$^{\dag}$}{E-mail address: \jjh}
\vskip 0.2in
\affil
Theory Group
Blackett Laboratory
Imperial College
London SW7 2BZ
UK
\vskip 0.5in
\centerline {\rm Preprint IC 92-93/26. April, 1993}
\vskip 0.2in
\centerline {\rm Submitted to {\sl Physical Review D}}.
\vskip 1.0in
\abstract
{
This paper is generally concerned with understanding how the
uncertainty principle arises in formulations of quantum mechanics,
such as the decoherent histories approach, whose central goal is the
assignment of probabilities to histories.  We first consider
histories characterized by position or momentum projections at two
moments of time.  Both exact and approximate (Gaussian) projections
are studied. Shannon's information is used as a measure of the
uncertainty expressed in the probabilities for these histories.  We
derive a number of inequalities in which the uncertainty principle
is expressed as a lower bound on the information of phase space
distributions derived from the probabilities for two-time histories.
We go on to consider histories characterized by position samplings
at $n$ moments of time.  We derive a lower bound on the information
of the joint probability for $n$ position samplings. Similar bounds
are derived for histories characterized by samplings of other
variables. All lower bounds on the information of histories have
the general form $\ln \left( V_H / V_S \right) $, where $V_H$ is a
volume element of history space, which we define, and $V_S$
is the volume of that space probed by the projections.
We thus obtain a concise and general form of the uncertainty principle
referring directly to the histories description of the system, and
making no reference to notions of phase space.
}
\endtopmatter

\head {\bf I. Introduction}

A quantum-mechanical history is defined by an initial quantum state
at some time $t_0$, and by a sequence of propositions at a succession of times
$t_1, t_2 \cdots t_n$. The initial state is represented by a density matrix
$\rho$. Each proposition is represented by a set of projection operators
$P_{\a}$.
These are positive hermitian operators that are both exclusive and
exhaustive:
$$
\eqalignno{
P_{\a} P_{\beta} & = \delta_{\a \beta} \ P_{\a},
&(1.1) \cr
\sum_{\a} P_{\a} & = 1
&(1.2) \cr }
$$
Evolution between each projection is described by the unitary evolution
operator, $e^{-\ih Ht}$. The probability for histories described in this
way is given by the expression,
$$
p(\a_1, \a_2, \cdots \a_n) = \Tr \left( P_{\a_n}^n(t_n)\cdots
P_{\a_1}^1(t_1)
\rho P_{\a_1}^1 (t_1) \cdots P_{\a_n}^n (t_n) \right)
\eqno(1.3)
$$
where
$$
P^k_{\a_k}(t_k) = e^{\ih H(t_k-t_0)} P^k_{\a_k} e^{-\ih H(t_k-t_0)}
\eqno(1.4)
$$

Eq.(1.3) is central to any formulation of quantum mechanics whose
aim is the assignment of probabilities to histories.
One particular such approach
is the decoherent histories approach
[\cite{gellmann,hartle,griffiths,omnes,dowker}].
In that approach, the central aim is to find, for a given
Hamiltonian and initial state, the sets of histories of closed
quantum systems for which the
probabilities (1.3) satisfy the so-called ``probability sum rules''.
Loosely, these are the rules obtained by demanding that the
probability of a composite history is the sum of the probabilities
of the more elementary histories of which it is comprised. An
example of such a sum rule (of which there are many) is,
$$
p( \cdots
\a_{k-1}, \a_{k+1} \cdots) = \sum_{\a_k} p ( \cdots \a_{k-1}, \a_k,
\a_{k+1} \cdots ) \eqno(1.5)
$$
Histories satisfying these rules are said to be ``consistent'', or
``decoherent'', and it
is solely in terms of such histories that predictions may be made.
Quantum-mechanical interference means that these rules are generally
not satisfied, and demonstrating consistency is typically
non-trivial.

The formula (1.3) also arises in a different context. It
is a concise summary of the Copenhagen approach to the quantum
mechanics of measured subsystems. It incorporates both the unitary
evolution of states together with the ``collapse of the wave
function'' incurred as a result of measurement by an external
agency, modeled by the projection operators [\cite{wigner,ABL}].

Irrespective of which interpretational scheme one is concerned with,
the mathematical properties of the  expression (1.3) are of
interest. This paper is concerned with exploring those properties.

Our particular concern is the question of how the
uncertainty principle arises in (1.3).
The usual form,
$$
\Delta p \Delta q \ge { \hbar \over 2}
\eqno(1.6)
$$
is a simple consequence of Fourier transform of the wave function of
the system at a fixed moment of time. However, in formulations that
give a central role to (1.3), the state of the system of the system
at a fixed moment of time does not enter in a fundamental way.
Instead, all physically meaningful notions must be expressed through
the probabilities (1.3). It therefore becomes an important issue to
understand how these probabilities recognize the uncertainty
principle.
It is not difficult to see that it will arise
as a limitation on the degree to which (1.3) may be peaked about a
particular history. This is because the probability (1.3) is a
distribution over quantities that are generally non-commuting, so
one would not expect it to become arbitrarily peaked. The aim of
this paper is to establish the detailed form this limitation takes.

As a measure of the degree to which (1.3) is peaked, we shall use
the Shannon information:
$$
I = \ - \sum_{\a_1 \cdots \a_n} \ p(\a_1, \cdots \a_n) \ln
p(\a_1 \cdots \a_n )
\eqno(1.7)
$$
This measure, for histories, is in many ways more natural
and easier to use
than the variances, employed in (1.6). We shall show that the
uncertainty principle generally arises as a lower bound on the
information (1.7). In particular, for the case in which the
alternatives $\a_k$ are discrete, the probabilities (1.3) have an
upper bound $p_{max}$, over all initial states $\rho$ and over
all possible values of the alternatives. If there is a restriction
on the degree to which (1.3) is peaked, as one would expect when the
projections do not commute,
then $p_{max} < 1$. The
information (1.7) then has a  non-trivial lower bound
$$
I \ge \ln \left( {1 \over p_{max} } \right)
\eqno(1.8)
$$

We begin in Section II with a brief review of some properties of
Shannon information.
We then go on in Section III to discuss information-theoretic
measures of uncertainty in quantum mechanics. We review earlier work
on information-theoretic versions of the uncertainty principle,
expressed in terms of the
state of the system at a fixed moment of time.

In Sections IV and V we discuss quantum-mechanical histories of the
form (1.3) characterized by position and/or momentum projections at
two moments of time. We consider the case of both exact and
approximate (Gaussian) projection operators. The general idea is to
use the two-time histories to derive imprecise samplings of phase
space, and then compute lower bounds on the information of the
quantum-mechanical phase space distributions. In regimes where they
are non-trivial, we find that all of the bounds have the approximate
form,
$$
I(K,X) \ge \ln \left( { 2 \pi \hbar \over \s_x \s_k } \right)
\eqno(1.9)
$$
where $I(K,X)$ is the information of the phase space distributions
and $\s_x \s_k $ is the volume of phase space probed by the projections.

In Section VI we go on to study histories characterized by position
samplings at $n$ moments of time. We show that the uncertainty
principle arises as a restriction on the information of the
approximate form,
$$
I(X_1, X_2, \cdots X_n ) \ \ge \ \ln \left( { V_H \over
\s_1 \s_2^2 \cdots \s_{n-1}^2 \s_n } \right)
\eqno(1.10)
$$
in the regime where it is non-trivial.
Here, $\s_i$ is the width of the position sampling $i$, and $V_H^{-1}$ is
a ``density of paths'' factor. We argue that $V_H$ thus has the
interpretation as the ``fundamental volume of history space'',
analogous to the factor of $2 \pi \hbar $ in (1.9). We derive a result
identical in form for histories characterized by other types of
projections. We thus obtain a form of the uncertainty principle
which is both concise and general, and is phrased entirely in the
language of histories, without reference to phase space.
We summarize and discuss in Section VII.

Some words are in order concerning the use of Shannon
information for the probabilities (1.3). Since the quantities
defined by (1.3) generally do not in fact satisfy the probability
sum rules, such as (1.5), they cannot strictly be regarded as
probabilities. Use of the Shannon information (1.7) therefore
requires some qualification.  Although they do not obey the
probability sum rules, the (candidate) probabilities (1.3) are
non-negative and normalized, and thus the information (1.7) is a
well-defined  quantity, and may be used as a measure of the degree
of spread of the candidate probability. The important point is that
at no stage are the probabilities sum
rules assumed, and thus no inconsistencies arise.

It is of course an interesting question, from the perspective of the
decoherent histories approach, to extend the considerations of the
present paper to the case in which the candidate probabilities (1.3)
do obey the probability sum rules. Decoherence may be achieved, for
example, by coupling the system of interest to an environment.
Modifications of the uncertainty relations (1.9), (1.10) due to
environmentally-induced ({\it e.g.} thermal) fluctations can then be
expected. This is considered in Refs.[\cite{ander,jjh4}]. The
information-theoretic inequalities considered here then become
conditions that such decoherering probabilities must satisfy in the
limit that the coupling to the environment goes to zero.

\head {\bf II. Information Theory}

In this section, we briefly review some results from information
theory.
This section solely concerns generic probability distributions, and makes
no reference to quantum mechanics.

Let $p_i$ be the probabilities
for a data set $S$ consisting of
discrete set of alternatives labeled by $i$, $i = 1,2 \cdots N $. One has $0
\le p_i \le 1$ and $ \sum_i p_i = 1 $. The {\it information}
of the data set $S$ is defined to be
$$
I(S) = - \sum_{i=1}^N p_i \ln p_i
\eqno(2.1)
$$
Here, $\ln$ is the logarithm to base $e$. $I(S)$ satisfies the inequalities
$$
0 \le I(S) \le \ln N
\eqno(2.2)
$$
It reaches its minimum if and only if $p_i = 1$, for one particular value of
$i$, and so $p_i = 0$ for all the other values. It reaches its maximum when
$p_i = {1 \over N} $ for all $i$. The information of a probability distribution
is therefore a measure of how strongly peaked it is about a given alternative.
For this reason, $I(S)$ is sometimes referred to as {\it uncertainty}, being
large for spread out distributions and small for concentrated ones. $I(S)$ is
sometimes also referred to as the {\it entropy} of the distribution, but we
shall not use that nomenclature here.

Base $2$ is often used in the definition (2.1). In this case $I(S)$
has the interpretation as the average number of bits required to specify an
alternative, given that alternative $i$ occurs with probability $p_i$.

Information may also be defined for
continuous probability distributions. Let $X$ be a random variable with
probability density $p(x)$. Then $\int dx \ p(x) = 1$. The information of
$X$ is defined to be
$$
I(X) = - \int dx \ p(x) \ln p(x)
\eqno(2.3)
$$
Unlike the discrete case, $I(X)$ is no longer positive, since $p(x)$
is not a probability, but a probability density, so may be greater than $1$.
However, it retains its utility as a measure of uncertainty. This is
exemplified by a Gaussian distribution of variance $\D x$,
$$
p(x) = { 1 \over \left( 2 \pi (\D x)^2 \right)^{\half} } \
\exp \left( - { (x - x_0)^2 \over 2 (\D x)^2 } \right)
\eqno(2.4)
$$
It has information
$$
I(X) = \ln \left( 2 \pi e (\D x)^2 \right)^{\half}
\eqno(2.5)
$$
{}From this we see that $I(X)$ may be unbounded from below, and indeed,
approaches $- \infty$ as $\D x \rightarrow 0$ and $p(x)$ approaches a
delta-function. $I(X)$ is also unbounded from above, as may be
seen by taking the width $\D x$ to be very large. However, if the variance
is fixed, then a straightforward variational calculation shows that $I(X)$ is
maximized by the Gaussian distribution (2.4). Eq.(2.5) therefore represents an
upper bound on the information of probability distributions with variance $\D
x$,
$$
I(X) \ \le \
 \ln \left( 2 \pi e (\D x)^2 \right)^{\half}
\eqno(2.6)
$$
with equality if and only if $p(x)$ is a  Gaussian.

The literature contains a vast number of results about information.
We will record only one, since it will be needed later.
Suppose from a probability distribution
$p(x)$ one constructs a ``coarser-grained'' probability distribution
$$
q(\bar x) = \int dx \ f(\bar x, x) \ p(x)
\eqno(2.7)
$$
for some smearing function $f(\x, x)$ satisfying
$ \int d \bar x  f(\bar x, x) = 1$. Then if we denote the information of
$q(\bar x)$ by $I(\bar X)$, it may be shown that
$$
I(\bar X) \ \ge \ I(X)
\eqno(2.8)
$$
This inequality expresses the intuitive idea that smearing or coarse-graining a
probability distribution increases the amount of uncertainty it expresses.
A corresponding result also holds for the discrete case. The result,
for both the continuous and discrete case,
follows readily from the convexity of the function $x \ln x $, so we
shall refer to this result as the convexity property.

For further details on information theory, see Refs.[\cite{shannon,cover}].

\head {\bf III. Information-Theoretic Uncertainty Relations}

We now describe a number of information-theoretic expressions of the
uncertainty principle. We begin by describing the projection operators
used to sample position and momentum.

\subhead {\bf III(A). Samplings of Position and Momentum}

Approximate samplings of position may be carried out using projection
operators. The projection operators effect a partition of the real line
into regions (or ``bins'') of size $\s_x$. Explicitly, they take the form
$$
P^x_{\a} = \int dx \ \U (x - \x_{\a}) \ |x \ra \la x |
\eqno(3.1)
$$
where $ \U(x-\x_{\a})$ is a sampling function. The most appropriate choice
is to take it to be
$$
\U(x-\x_{\a})= \theta \left({x-\x_{\a}+ \half \sigma_x \over \s_x } \right) \
 \theta \left( {-x + \x_{\a}+ \half \sigma_x \over \s_x }\right)
\eqno(3.2)
$$
It is equal to $1$ in an interval of size $\sigma_x$ centred around $\x_{\a}$
and zero otherwise, where $\x_{\a} = \a \s_x $, and $\a$ is an integer.
We will generally use a bar to denote coarse-grained variables.
The sampling function satisfies the relations,
$$
\eqalignno{
\sum_{\a} \ \U (x - \x_{\a} ) &= 1
&(3.3) \cr
\int dx \ \U (x - \x_{\a} ) &= \s_x
&(3.4) \cr }
$$
Eq.(3.3) ensures that the projections are exhaustive.  They are exclusive
because $\U$ vanishes outside a unit interval.

Another
choice for $\U$ which is sometimes convenient is a Gaussian of width
$\sigma_x$,
$$
\U(x-\x_{\a}) = {1 \over ( 2 \pi )^{\half} }
\ \exp \left( - { (x-\x_{\a})^2 \over 2 \sigma_x^2 } \right)
\eqno(3.5)
$$
Again $\x_{\a} = \a \s_x $, but $\a$ is now a continuous label. The properties
(3.3) and (3.4) still hold, given the convention that the summation over
$\a$ is now an integration. With this choice of $\U$ the projections are
only approximately exclusive. This means that the label $\a$,
although continuous, really has significance only up to order 1.

The case of precise samplings, $P_{\x} = | \x \ra \la \x |$, is obtained by
writing $P_{\x} = \s^{-1} P_{\a}^x $, and
letting $\s_x \rightarrow 0 $, and one has
$$
\s_x^{-1} \U (x- \x_{\a}) \rightarrow
\delta (x-\x), \quad \quad \s_x \sum_{\a} \rightarrow \int d \x
\eqno(3.6)
$$
In a similar manner, one may construct projections for samplings of  momentum,
$$
P^k_{\b} = \int dk \ \G(k-\k_{\b}) \ |k \ra \la k |
\eqno(3.7)
$$
for some sampling function $ \G(k-\k_{\b}) $, of width $\s_k$, where
$\k_{\b} = \b \s_k $.

\subhead{\bf III(B). Samplings of Two Ensembles}

The first result we shall describe envisages a situation in which one
has two ensembles, prepared in an identical state.
Samplings of position are made on the first ensemble,
and samplings of momentum are made on the second.

Consider a position sampling of a system described by a density matrix
$\rho$. The probability that the result lies in the region labeled by
$\a$ is,
$$
\eqalignno{
p^x(\a) &= \Tr \left( P_{\a}^x \rho \right) \cr
&= \int dx \ \U (x-\x_{\a}) \ \la x | \rho | x \ra
&(3.8) \cr }
$$
We wish to use the information $I_{\rho} (\bar X)$ as a measure of uncertainty
in the probability distribution $p^x(\a)$. By the convexity
property (2.8), one has
$$
\eqalignno{
I_{\rho} (\bar X) \  & \equiv \ - \sum_{\a} p^x(\a) \ln p^x (\a) \cr
& \ge \ - \int dx \ \la x | \rho | x \ra \ln \la x | \rho |
x \ra  - \ln \s_x
\cr &
\equiv I_{\rho}(X) - \ln \s_x
&(3.9) \cr }
$$
The $\ln \s_x $ term arises because of (3.4)

In a similar manner, we can consider a momentum sampling on an identically
prepared system, giving the probability distribution
$$
p^k(\b) = \Tr \left( P_{\b}^k \rho \right)
\eqno(3.10)
$$
One may compute its information, $ I_{\rho}(\bar K)$, and one has,
$$
I_{\rho}(\bar K) \ge I_\rho (K) - \ln \s_k
\eqno(3.11)
$$

A general density operator $\rho$ may be written
$$
\rho = \sum_i \ c_i \ | \psi_i \ra \la \psi_i |
\eqno(3.12)
$$
for some set of states $|\psi_i \ra $. Again using the
convexity property (2.8), one has
$$
I_{\rho} (X) \ \ge \ \sum_i \ c_i \ I_{\psi_i} (X)
\eqno(3.13)
$$
where $ I_{\psi_i} (X) $ denotes the information of the probability
distribution obtained from precise sampling of the pure state $|\psi_i \ra $.
A similar result holds for momentum samplings.
It follows that there exists a pure state $|\psi \ra $
such that
$$
I_{\rho}(\bar X)+ I_{\rho}(\bar K) \ \ge \ I_{\psi} (X) + I_{\psi} (K)
- \ln ( \s_x \s_k )
\eqno(3.14)
$$
with equality for precise samplings and $\rho = | \psi \ra \la \psi | $.
(Note that the $\ln (\s_x \s_k)$ term dissappears in the limit of
precise samplings, since it is taken into the projections on the left-hand
side, as described above).

$I_{\psi}( K)$ and $I_{\psi}(X)$ are individually unbounded from
below, since one can always find states which are arbitrarily peaked in either
position or in momentum. However, a state strongly peaked in position, and
hence with large negative $I_{\psi}(X)$, will be very spread out in
momentum, and thus $I_{\psi} (K)$ will be large and positive. It is
therefore plausible that the uncertainty principle will express
itself as a lower bound on the sum, $I_{\psi}(X)+ I_{\psi}(K)$.
The usual inequality expressing the uncertainty principle,
$$
\D x \ \D k  \ \ge \ { \hbar \over 2}
\eqno(3.15)
$$
achieves equality for the minimum uncertainty wave packets. Since they
are Gaussians, we immediately have, from (2.5),
$$
\eqalignno{
I_{\psi} (X) + I_{\psi} (K) &= \ln \left( 2 \pi e \D x \D k \right) \cr
& = \ln ( \pi e \hbar )
&(3.16) \cr }
$$
for the minimum uncertainty wave packets (coherent states). It was therefore
conjectured
by Everett [\cite{everett}] that the uncertainty principle may be expressed in
information-theoretic terms as
$$
I_{\psi} (X) + I_{\psi} (K) \ \ge \  \ln ( \pi e \hbar)
\eqno(3.17)
$$
He also noted that this inequality implies the usual form of the
uncertainty principle. To see this, recall from Section II that the information
of a probability distribution is bounded from above by the information of a
Gaussian of the same variance. This means that
for any state $|\psi \ra $, with variances $\D x$, $\D k$, one has
$$
\ln \left( 2 \pi e \D x \D k \right) \ \ge
\ I_{\psi} (X) + I_{\psi} (K)
\eqno(3.18)
$$
The usual uncertainty principle then follows immediately by comparing
(3.17) and (3.18).
The inequality (3.17) was proved by Beckner [\cite{beckner}], Bialynicki-Birula
and Mycielski [\cite{bialy2}] and Hirschmann [\cite{hirsch}],
using the Hausdorff-Young inequalities from Fourier analysis.

Combining all of the above results, we have
$$
I_{\rho} (\bar X) + I_{\rho} (\bar K) \ \ge \  1+ \ln ( { \pi  \hbar \over
\s_x \s_k } )
\eqno(3.19)
$$
with equality for precise samplings and $\rho$ a minimum uncertainty
wavepacket. Eq.(3.19) represents a very modest generalization of (3.17)
to the case of imprecise samplings of position and momentum.

\subhead {\bf III(C). Samplings of a Single Ensemble}

Of greater interest for our purposes
is the situation in which the samplings of
position and momentum are made on the {\it same} system. There are a
number of ways of doing this, and we shall consider them in turn.
Perhaps the simplest is to carry out simultaneous but imprecise samplings
of both position and momentum. These may be effected using the coherent
state projectors, which we now describe.

The (canonical) coherent states [\cite{klauder}] may be defined to be the
states
$$
|z \ra = |p,q \ra = U(p,q) | 0 \ra
\eqno(3.20)
$$
where $ |0 \ra $ is the ground state of the harmonic oscicallator.
$U(p,q)$ is the unitary Weyl operator,
$$
U(p,q) = \exp \left( {i \over \hbar} (p \hat Q - q \hat P) \right)
\eqno(3.21)
$$
where $\hat Q$ and $\hat P$ denote the position and momentum operators.
In the position representation the coherent states are given by
$$
\la x | p,q \ra = {1 \over (2\pi \s_q^2)^{1/4} } \left( - {(x-q)^2 \over 4
\s_q^2 }  + \ih px \right)
\eqno(3.22)
$$
Their most important property is the completeness relation,
$$
\int { dp dq \over 2 \pi \hbar } \ | p,q \ra \la  p, q | = 1
\eqno(3.23)
$$
They are however only approximately orthogonal,
$$
\la p,q | p', q' \ra = \exp \left( {i \over 2\hbar } (p'q-q'p) - {1 \over 4}
\left[ {(p-p')^2 \over \s_p^2}+ {(q-q')^2 \over \s_q^2 }\right] \right)
\eqno(3.24)
$$
where $\s_p \s_q = {\hbar \over 2} $.
These properties suggest that we may regard the operator
$$
P_z = |p,q \ra \la  p, q |
\eqno(3.25)
$$
as an approximate projection operator affecting approximate simultaneous
samplings of position and momentum. The approximate orthogonality
property (3.24), means that the labels $p$ and $q$ are coarse-grained
momentum and position, having significance only up to
the widths $\s_p$ and $\s_q$ respectively.

If the state of the system is
described by a density operator $\rho$, the probability distribution
of approximate position $\x$ and approximate momentum $\k$ is therefore
$$
p(\k, \x) = \Tr ( P_z  \rho ) \ = \ \la \k, \x | \rho |
\k, \x \ra
\eqno(3.26)
$$
This probability is normalized in the measure $d\k d\x / 2 \pi \hbar $.
Consider the information of this distribution,
$$
I_{\rho} ( K, X) = - \int  { d\k d \x \over 2 \pi \hbar}
\ p(\k, \x) \ln p(\k, \x)
\eqno(3.27)
$$
If $ p(\k, \x) $ were a classical phase space distribution, then (3.27)
would be the usual entropy in statistical mechanics. The entropy would
be unbounded from below because in classical mechanics, the phase space
distribution may be arbitrarily concentrated about a particular region of
phase space. In quantum mechanics, by contrast, phase space distributions
concentrated on regions smaller in size than $\hbar$ would violate the
uncertainty principle. We therefore expect a lower bound on (3.27).

A reasonable guess as to what this lower bound should be is obtained by
evaluating (3.27) with $\rho$ a coherent state, since the coherent states are
normally thought of as being the states most concentrated in phase space.
Writing $\rho = |z \ra \la z | $, one finds
$$
I_{|z \ra } ( K, X) = 1
\eqno(3.28)
$$
For these reasons, it was conjectured by Wehrl [\cite{wehrl}] that
$$
I_{\rho} ( K,  X) \ \ge \ 1
\eqno(3.29)
$$
with equality if and only if $\rho$ is a coherent state. This was subsequently
proved by Lieb [\cite{lieb}], again using some inequalities from Fourier
analysis
(best constants in the Hausdorff-Young and Young inequalities).

A simple but important generalization of this result was noted by Grabowksi
[\cite{grabowski}]. This is that the inequality (3.29) continues to hold for
projections
constructed from a class of {\it generalized} coherent states,
namely, those of the form
$$
| \psi_{\x\k} \ra = U(\x,\k) | \psi \ra
\eqno(3.30)
$$
where $ | \psi \ra $ is an arbitrary state. The point is that they share
with the usual coherent states the completeness relation, (3.23), and
it is this property that is exploited in Lieb's proof.

This generalization also permits a connection with the usual uncertainty
relation to be made. One has
$$
\eqalignno{
I_{\rho}(K,X) \ & \le \ I_{\rho}(X) + I_{\rho}(K) \cr
& \le \ 1 + \ln \left( { \Delta x \Delta k \over \hbar } \right)
&(3.31) \cr }
$$
where $ \Delta x $ and $ \Delta k $ are the variances of $x$ and $k$ in the
probability distribution (3.26), but with the generalized coherent states
(3.30). The first inequality is a standard property
of information; the second is the inequality (2.6) used twice
(up to a factor
of $ 2 \pi \hbar $, because of our choice of phase space measure).
Together with (3.24), (3.31) implies that
$$
\Delta x \Delta k \ \ge \ \hbar
\eqno(3.32)
$$
This is not the usual uncertainty relation (no factor of $\half$), because
the variances express not only the uncertainty in the initial state, but
also the uncertainty in the projections, which are imprecise. Indeed,
one has
$$
\eqalignno{
(\Delta x)^2 &= (\Delta_{\rho}  x)^2  + (\Delta_{\psi} x)^2
&(3.33) \cr
(\Delta k)^2 &= (\Delta_{\rho}  k)^2  + (\Delta_{\psi} k)^2
&(3.34) \cr }
$$
The first term on the right-hand side of each relation
is the variance in the initial state;
the second is the variance in the generalized coherent state projection
with fiducial state $ |\psi\ra $. Choosing $\rho$ to be the pure state
$ | \psi \ra \la \psi | $, one thus obtains the usual uncertainty relation,
$$
\Delta_{\psi} x \Delta_{\psi} k \ \ge \ {\hbar \over 2 }
\eqno(3.35)
$$
An alternative method of connecting (3.29) with the usual
uncertainty relations may be found in Ref.[\cite{ander}].

Results similar to (3.17) and (3.19) have been obtained  by Deutsch
[\cite{deutsch}], Partovi [\cite{partovi1}]. and Maassen and Uffink
[\cite{maassen}]. Eq.(3.17) has been generalized to include thermal
fluctuations at thermal equilibrium by Abe and Suzuki [\cite{abe}].
Anderson and Halliwell have generalized (3.29) to include thermal
fluctuations in a class of non-equilibrium systems [\cite{ander}]
(see also Ref.[\cite{jjh4}]). For an alternative approach to unsharp
 samplings of non-commuting observables using positive
operator-valued measures, see Schroeck [\cite{schroeck}]. For other
related results on information-theoretic uncertainty relations, not
directly relevant to the present paper, see
Refs.[\cite{bialy1,busch,grabowski2,mamojka,partovi2,raja,sanchez,cover2}].

\head {\bf IV.  Two-Time Histories -- Approximate Projectors}

We now show how to obtain information-theoretic uncertainty
relations for histories characterized by projections at two moments
of time. The projections will be onto position at two moments of
time, or onto momentum and position. The important feature is that
the time-dependent projections $P^k_\a(t_k)$ do not commute, and so
one would not expect their probability distributions to be
arbitrarily peaked. We therefore expect to derive lower bounds on
the information, in analogy with (3.29).

The case of position and momentum samplings by exact projections, such
as Eq.(3.2), is quite different from the case of approximate
projections, such as Eq.(3.5), and each case needs to be treated
separately. The approximate projection case is a direct extension of
the results of Section III(C), and we consider this case first. The
case of exact projections will be treated in the next section.

\subhead{\bf IV(A). A Lower Bound on the Information}

In brief, the idea is as follows. The probabilities for histories are
most generally given by an expression of the form
$$
p({\bf \au}) = \Tr ( C_{{\bf \au}}^{\dag} C_{{\bf \au}} \rho )
\eqno(4.1)
$$
where $C_{{\bf \au}}$ denotes a string of time-dependent projection
operators,
$$
C_{\au} = P_{\a_n}^n (t_n) \cdots P_{\a_1}^1 (t_1)
$$
and we use the notation ${\bf \au}$ to denote a string of $\a$'s.
The burden of the results described
below will be to show that for the case of two-time histories
considered here, the operator $ C_{{\bf \au}}^{\dag} C_{{\bf \au}} $
may be written in the form,
$$
C_{{\bf \au}}^{\dag} C_{{\bf \au}} = U(\k,\x) \ \Omega  \ U^{\dag} (\k, \x)
\eqno(4.2)
$$
for some operator $\Omega$. The point is that
the dependence on the sampled positions and momenta $\x$ and $\k$
resides entirely in the unitary Weyl operator
$U(\k, \x) $. Now $ C_{\au}^{\dag} C_{\au} $ is a positive hermitian operator,
and thus $\Omega$ is also. It may therefore be written
$$
\Omega = \sum_a \lambda_a \ | a \ra \la  a |
\eqno(4.3)
$$
where the coefficients $\lambda_a$ are positive. The probabilities
(4.1) for our two-time histores may now be written
$$
p(\a,\b) = \sum_a \ \lambda_a \ \la a | \ U(\k,\x) \ \rho \ U^{\dag} (\k,\x)
 \ | a \ra
\eqno(4.4)
$$
Here, we have introduced, as earlier, the continuous bin labels $\a$ and $\b$,
defined in terms of the sampled positions and momenta
by $\x = \a \s_x $, $\k = \b \s_k $, where
$ \s_x $ and $\s_k $ are their respective widths.
The right-hand side of (4.4) involves the expectation value of
$\rho$ in the generalized
coherent states, $ | a_{\k\x} \ra  = U^{\dag}(\k,\x)  | a \ra $. By
the convexity property (2.8), the information of (4.4) satisfies
the inequality,
$$
\eqalignno{
I(K,X) \ &  \equiv - \int d \a d \b \ p(\a,\b) \ln p(\a,\b)
\cr
& \ge \ - \int d\x d\k \ \la  a_{\k\x} | \rho | a_{\k\x} \ra
\ln \ \la  a_{\k\x} | \rho | a_{\k\x} \ra  - \ln(\s_x \s_k)
&(4.5) \cr }
$$
The factor of $\ln (\s_x \s_k)$ arises from the change of
variables from $\a,\b$ to $\x,\k$.
{}From the previous section, Eqs.(3.29), (3.30), we thus deduce the inequality,
$$
I(K,X) \ \ge \ 1 + \ln \left( {2 \pi \hbar \over \s_x \s_k } \right)
\eqno(4.6)
$$
The factor of $2 \pi \hbar $ appears because of the difference in
phase space measures used in (3.27) and (4.5). The factor of $2$
difference between (4.6) and (3.19) is due to the fact that at
equality, (3.19) measure the uncertainty in the state alone, whereas
(4.6) also includes the uncertainty in the coherent state projector.

Eq.(4.6) is an intuitively appealing result. The argument of the
logarithm is the inverse of
the number of elementary cells of phase space sampled.
If that number is large, {\it i.e.},
$\s_x \s_k >> 2 \pi \hbar $, then the lower
bound approaches $- \infty$, and thus the uncertainty principle
imposes little restriction on samplings of phase space large
compared to the fundamental cell. On the other hand, the bound becomes
significant when $\s_x \s_k $ is of order $ 2 \pi \hbar $ or
smaller, in agreement with the expectation that the uncertainty
principle imposes limitations on samplings comparable to the
size of the fundamental cell.

Everything up to Eq.(4.5) is also true for the discrete case (with
the integral over $\a$, $\b$, replaced by a discrete sum), but it
is not possible to deduce the inequality (4.6), since this holds
only for the continous case.

We now need to show that the projections satisfy the condition (4.2)
for the two-time histories of interest. It is also necessary to calculate
$\Omega$, to determine the conditions under which the inequality becomes
equality. Before that, we need to describe some mathematical tools.

%

\subhead {\bf IV(B). The Weyl Calculus}

The analysis of (4.1) is conveniently carried out with the aid
of a set of mathematical tools referred to as the Weyl calculus. This in turn
is part of a larger area of mathematics called microlocal analysis
[\cite{omnes2}].
The basic
idea is to define a one-to-one correspondence between every self-adjoint
operator, $\hat A$ say, on the Hilbert space, and a real function $A(p,q)$
defined in a phase space, referred to as the Weyl symbol of $\hat A$.
A particular example of how this correspondence may be obtained is through
the Wigner transform,
$$
W_A(p,q) = {1 \over 2 \pi \hbar} \int_{-\infty}^{\infty} \ d \xi \
\ \la q+ {\xi \over 2} | \hat A | q - {\xi \over 2}  \ra \ e^{-{i \over \hbar}
p \xi}
\eqno(4.7)
$$
When the operator $ \hat A $ is the density operator, $ \hat \rho $,
$W_{\rho}$ is called the Wigner function. It shares many properties
of classical phase space distributions, although it is often not positive.
It has been used extensively in discussions of the classical limit
[\cite{gellmann,anderson,wignerfn,jjh}].
We shall make use of the Wigner transform (4.7) to analyse (4.1).

An alternative form of (4.7) that we shall find more useful is,
$$
W_A(p,q)  = \Tr \left( \hat \Delta (p,q) \hat A \right)
\eqno(4.8)
$$
where
$$
\hat \Delta (p,q) = {\hbar \over 2 \pi} \int du dv \ e^{ipu + iqv}
\ e^{-iu \hat P -iv \hat Q}
\eqno(4.9)
$$
Here, $\hat Q$ and $\hat P$ are the usual position and momentum operators,
satisfying $ [ \hat Q, \hat P] = i \hbar $.

We record and prove some useful properites of the Wigner transform.
First, one has
$$
\Tr ( \hat A \hat B ) = { 1 \over 2 \pi \hbar } \int dp dq \ W_A(p,q) W_B(p,q)
\eqno(4.10)
$$
This follows readily from inserting the explicit form (4.7) into the
right-hand side of (4.10).

Next, we discuss the properties of the Weyl symbol under shifts of
its arguments. Introduce the unitary Weyl operator,
$$
U(p, q) = e^{\ih p \hat Q - \ih q \hat P}
\eqno(4.11)
$$
It has the properties,
$$
\eqalignno{
U^{\dag}(\bar p, \bar q) \hat Q U(\bar p, \bar q) &= \hat Q + \bar q
&(4.12) \cr
U^{\dag}(\bar p, \bar q) \hat P U(\bar p, \bar q) &= \hat P + \bar p
&(4.13) \cr }
$$
The Baker-Campbell-Hausdorff relation is
$$
e^{\hat A + \hat B} = e^{\hat A} e^{\hat B} e^{\half [\hat A, \hat B]}
\eqno(4.14)
$$
if $[\hat A, \hat B]$ commutes with $\hat A$ and $\hat B$. It follows that,
$$
U^{\dag}(\bar p, \bar q) \ e^{-iu \hat P -iv \hat Q} \ U(\bar p, \bar q)
= \ \ e^{i\bar pu + i\bar qv} \ e^{-iu \hat P -iv \hat Q}
\eqno(4.15)
$$
and thus
$$
U^{\dag}(\bar p, \bar q) \ \hat \Delta(p,q) \ U(\bar p, \bar q)
= \hat \Delta(p+\bar p,q+ \bar q)
\eqno(4.16)
$$
{}From this we see that
$$
\eqalignno{
W_A(p+\bar p,q+\bar q)  &= \Tr \left( \hat \Delta (p+\bar p ,q+\bar q)
\hat A \right)
\cr &= \Tr \left( \hat \Delta (p,q) {\hat A}' \right)
\cr &= W_{A'}(p,q)
&(4.17) \cr }
$$
where ${\hat A}' = U(\bar p, \bar q) \hat A U^{\dag} (\bar p, \bar q) $.
That is, translating the coordinates and momenta of the Weyl symbol
are equivalent to a unitary transformation under the Weyl operator of the
original operator.

{}From (4.1) and (4.10), it follows that the probabilities for histories
characterized by the chain operator $C_{\au}$ are given by
$$
p(\au) = {1 \over 2 \pi \hbar} \int dp dq \ W_{C^{\dag} C}(p,q) \ W_{\rho}
(p,q)
\eqno(4.18)
$$

\subhead {\bf IV(C). Position and Direct Momentum Samplings}

The first type of history we shall consider is one characterized by
an imprecise position sampling at time zero and an imprecise momentum
sampling at time $t$. The probability for this history is given by
$$
p(\a, \b, t ) = \Tr \left[ P_{\b}^k e^{-iHt} P_{\a}^x \rho P_{\a}^x
e^{iHt} \right]
\eqno(4.19)
$$
In the short time limit, employed here, evolution is described by
the free Hamiltonian. This clearly commutes with the momentum
projections, and thus
$t$ drops out in the short time limit. One thus has
$$
C_{\au}^{\dag} C_{\au} = P_{\a}^x P_{\b}^k P_{\a}^x
\eqno(4.20)
$$
The Weyl symbol of this operator is
$$
W_{C^{\dag} C}(p,q) = {1 \over 2 \pi \hbar} \int \ d \xi
\ \la q+ {\xi \over 2} |P_{\a}^x P_{\b}^k P_{\a}^x
| q - {\xi \over 2}  \ra \ e^{-{i \over \hbar} p \xi}
\eqno(4.21)
$$
Inserting the explicit forms for the projection operators, one obtains,
$$
W_{C^{\dag} C}(p,q) = {1 \over 2 \pi \hbar} \int \ d \xi dk
\ e^{\ih \xi(k-p)} \ \U (q+ \half \xi - \x) \ \U ( q - \half \xi -\x)
\G (k - \k)
\eqno(4.22)
$$
Letting $k \rightarrow k + \k $, it is readily seen that one has,
$$
W_{C^{\dag} C}(p,q) = W_{\Omega}(p-\k,q-\x)
\eqno(4.23)
$$
Here, $\Omega$ is the operator whose Weyl symbol is (4.22), but with
$\k=0$ and $\x=0$. $\Omega$ is therefore equal to $ P_{\a}^x P_{\b}^k
P_{\a}^x $ at $\k=0$ and $\x=0$, that is,
$$
\Omega = {1 \over 2 \pi \hbar} \int dx dy dk \ \U (x) \U ( y) \G (k)
\ e^{\ih k(x-y)}
\ |x \ra \la y |
\eqno(4.24)
$$
{}From (4.17), (4.23), we therefore have a result of the general
form (4.2),
$$
P_{\a}^x P_{\b}^k P_{\a}^x =
U^{\dag}(\bar k, \bar x) \ \Omega \ U(\bar k, \bar x)
\eqno(4.25)
$$
{}From the above, it therefore follows that the information of the phase space
distribution (4.19) obeys the inequality (4.6).

Consider now the conditions for equality. Equality is obtained if and only
if both $\rho$ and $\Omega$ are of the form $|z\ra \la z|$, where $|z \ra $
is a canonical coherent state, (3.20). From (4.24), one can see that
$\Omega$ will be of that form, if and only if $\G (k) = \delta (k) $
and $\U(x)$ is a Gaussian. That is, the first projection is a Gaussian
projection onto position, and the second is an infinitely precise
sampling of momentum.

\subhead {\bf IV(D). Position and Time-of-Flight Momentum Samplings}

We now consider a history characterized by imprecise position
samplings at times $0$ and $t$.
The probability is
$$
p(\a_1, \a_2, t ) = \Tr \left[ P_{\a_2}^x e^{-iHt} P_{\a_1}^x \rho
P_{\a_1}^x e^{iHt} \right]
\eqno(4.26)
$$
{}From this one can construct a phase space probability $p(\a_1, \b,
t)$, where $ \b \s_k = \k$, and $\k = m(\x_2 - \x_1)/t$, for small
$t$. We have $\x_1 = \s_x \a_1 $, $\x_2 = \s_x \a_2$, thus $\b =
\a_2 - \a_1$ and $\s_k = m \s_x / t $. One has,
$$
C_{\au}^{\dag} C_{\au} = P_{\a_1}^x e^{iHt} P_{\a_2}^x  e^{-iHt} P_{\a_1}^x
\eqno(4.27)
$$

We will analyse this case for small times $t$. It is readily shown that
the Weyl symbol is
$$
\eqalignno{
W_{C^{\dag} C}(p,q) = & {1 \over 2 \pi \hbar} \int \ d \xi dx_2
\ e^{-\ih p \xi } \ \U (q+ \half \xi - \x_1) \ \U ( q - \half \xi -\x_1)
\ \U( x_2 - \x_2 )
\cr & \times
\la x_2, t | q - \half \xi, 0 \ra
\ \la x_2, t | q + \half \xi, 0 \ra^*
&(4.28) \cr }
$$
Now in the short time limit, the propagator is given by
$$
\la x_2, t | q - \half \xi, 0 \ra
= \left( { m \over 2 \pi \hbar i t } \right)^{\half}
\ \exp \left( {im \over 2\hbar t} (x_2 - q + \half \xi )^2 \right)
\eqno(4.29)
$$
Inserting this in (4.28), and performing the shift
$ x_2 \rightarrow x_2 + \x_2 $, one finds that the answer may be written
in the form
$$
\eqalignno{
W_{C^{\dag} C}(p,q) & =  \ { 1 \over 2 \pi \hbar} \int \ d \xi dx_2
\ \U (q+ \half \xi - \x_1) \ \U ( q - \half \xi -\x_1) \ \U( x_2 )
\ { m \over 2 \pi \hbar t }
\cr & \times
\ \exp \left( -\ih  \left[ p - {m \over t}(\x_2 -\x_1) \right] \xi
+ i {m \over \hbar t} ( x_2 - q + \x_1 ) \xi \right)
&(4.30) \cr }
$$
One therefore has
$$
W_{C^{\dag} C}(p,q) = W_{\Omega}( p - \k, q - \x_1)
\eqno(4.31)
$$
$\Omega$ is the operator whose Wigner
transform is (4.30) with $\x_1=0$ and $\x_2 = 0 $. Explicitly,
$$
\Omega = \int dx dy dx_2 \ \la x_2, t | y, 0 \ra \ \la x_2, t | x, 0 \ra^*
\ \U(x) \U(y) \U(x_2) \ |x \ra \la y |
\eqno(4.32)
$$
{}From (4.17), (4.31), we now have the result,
$$
P_{\a_1}^x e^{iHt} P_{\a_2}^x  e^{-iHt} P_{\a_1}^x
= U^{\dag}(\bar k, \bar x) \ \Omega \ U(\bar k, \bar x)
\eqno(4.33)
$$
We therefore again have the inequality (4.6), for the information of
the phase space distribution constructed from (4.26).

Consider the conditions for equality. Again this is achieved when
both $\rho$ and $\Omega$ are of the form $|z\ra \la z |$. This means
that Eq.(4.30) must be the Wigner transform of a coherent state,
{\it i.e.}, a product of Gaussians in $p$ and $q$. This can be
achieved by letting the width of the sampling function at $t$ go to
a delta-function, setting the sampling function at $t=0$ to a
Gaussian, and then letting $ t \ria \infty $. (This may be seen
explicitly in Ref.[\cite{jjh}]). We are, however, working in
the short time approximation, so this procedure can be carried out
only for the free particle case, for which the short time
approximation is exact.

It is also possible to deduce a lower bound on the information of
the joint probability for position samplings, (4.26). The
information of $p(\a_1, \a_2,t)$ is in fact equal to that of
$p(\a_1, \b, t)$, because the Jacobean of the transformation between
these variables is unity. One thus has the following bound on the
information of (4.26):
$$
\eqalignno{
I(X_1, X_2) \ \equiv & \ - \int d \a_1 d \a_2 \ p(\a_1, \a_2) \ln
p(\a_1, \a_2)
\cr
\ge & \ 1 + \ln \left( {2 \pi \hbar t \over m \s_x^2}
\right)
&(4.34) }
$$
This is strictly speaking a trivial rewriting of (4.6). We
record the result because it will be generalized to an arbitrary
number of position samplings in Section VI.

\subhead{\bf IV(E). Position Samplings at Arbitrary Time Separations}

The previous case concerned position samplings for any Hamiltonian, but
in the limit of small time separations.  For the case of linear
systems, we may extend this analysis to arbitrary time separations.
We now outline how this is done.

The propagator for linear systems is given by,
$$
\la x^{\pp},t^{\pp} | x',t' \ra = \Delta(t^{\pp},t')
\ \exp \left( \ih S(x^{\pp},t^{\pp}|x',t') \right)
\eqno(4.35)
$$
where $S$ is the action of the classical solution connecting initial
and final points, and is quadratic in the $x$'s. The prefactor
$\Delta$ is independent of the $x$'s, and is given by
$$
\Delta(t^{\pp},t') = \left[ - {1 \over 2 \pi i \hbar} { \partial^2
S(x^{\pp},t^{\pp}| x',t') \over \partial x^{\pp} \partial x'}
\right]^{\half}
\eqno(4.36)
$$
Repeating the analysis of the previous subsection,
Eq.(4.28) thus has the form
$$
\eqalignno{
W_{C^{\dag} C}(p,q) & =  \ { 1 \over 2 \pi \hbar} \int \ d \xi dx_2
\ \U (q+ \half \xi - \x_1) \ \U ( q - \half \xi -\x_1) \ \U( x_2 -
\x_2 ) \ | \Delta |^2
\cr & \times
\ \exp \left( -\ih p \xi
+ \ih S(x_2,t| q- \half \xi,0) - \ih S(x_2, t | q +
\half \xi, 0 ) \right)
&(4.37) \cr }
$$
Now, letting $x_2 \ria x_2 + \x_2$, and using the fact that $S$ is
quadratic, (4.37) may be written,
$$
\eqalignno{
W_{C^{\dag} C}(p,q) & =  \ { 1 \over 2 \pi \hbar} \int \ d \xi dx_2
\ \U (q+ \half \xi - \x_1) \ \U ( q - \half \xi -\x_1) \ \U( x_2 )
\ | \Delta |^2
\cr & \times
\ \exp \left( -\ih \xi ( p -\k)
- \ih \xi { \partial S \over \partial q} (x_2,t | q - \x_1, 0 )
\right)
&(4.38) \cr }
$$
where we have introduced
$$
\k = - { \partial S \over \partial \x_1} (\x_2, t | \x_1, 0 )
\eqno(4.39)
$$
{}From Hamilton-Jacobi theory, $\k$ is the initial momentum for the
classical path between $\x_1$ and $\x_2$. Now the point is that
(4.38) depends on $\x_2$ and $\x_1$ only through the combinations
$p- \k$ and $q- \x_1$,
and we again have a result of the form (4.33), but this time with
$\k$ given by (4.39). We therefore again deduce the inequality
(4.6), for the information of the corresponding phase space distribution.

What is perhaps more interesting in this case is to derive the
generalization of (4.34). Since $\k$ is linear in $\x_2$, $\x_1$ in
(4.39), we have,
$$
\k = { \partial \k \over \partial \x_2 } \x_2 + { \partial \k \over
\partial \x_1}
\x_1
\eqno(4.40)
$$
and thus,
$$
\b = { \s_x \over \s_k } \left(
 { \partial \k \over \partial \x_2 } \a_2 +
{ \partial \k \over \partial \x_1} \a_1 \right)
\eqno(4.41)
$$
Here, as before,
$\b \s_k = \k$, $\a_1 \s_x = \x_1$  and $\a_2 \s_x =
\x_2 $. Unlike the case of short times, the transformation from
$ \a_1, \b $ to $\a_1, \a_2 $ has non-trivial Jacobean. It follows
that
$$
I(X_1, X_2) = I(K,X)  - \ln \left( { \s_x \over \s_k }
\ \Biggl| { \partial \k \over \partial \x_2 } \Biggr| \right)
\eqno(4.42)
$$
Finally, using the bound (4.6) on $I(K,X)$, and noting that
$$
{ \partial \k \over \partial \x_2 } =
- { \partial^2 S \over \partial \x_1 \partial \x_2 }
(\x_2, t | \x_1, 0 )
\eqno(4.43)
$$
we derive the following bound on the information of position
samplings at arbitrary time separations,
$$
I(X_1,X_2) \ \ge \ 1 \ +
\ \ln \left( { 2 \pi \hbar \over \s_x^2 }
\ \Biggl| { \partial^2 S \over \partial \x_1 \partial \x_2 }
\Biggr|^{-1}
\right)
\eqno(4.44)
$$
We will generalize this result, and discuss it further in Section VI.

\head {\bf V. Two-Time Histories -- Exact Projectors}

As stated in the previous section, the case of exact projections is
rather different to the case of approximate ones and needs to be
treated separately. In this section we show how this is done.

We are again interested in an expression for the probability of a
two-time history of the form (4.1), where $C_{\au}^{\dag} C_{\au} $
is of the form (4.20) or (4.27). In each case it is again possible
to show that  $C_{\au}^{\dag} C_{\au} $ may be written in the form
(4.2), although note that now $\x_{\a}, \k_{\b}$ are discrete rather than
continuous variables. We can go on to use the steps (4.3) to (4.5),
except that the integral in (4.5) becomes a discrete sum, and it is
at this point that we can go no further. Of course, if the bin
sizes $\s_x$, $\s_k$ are very small, then the discrete sum may be
approximated by the continous integral (4.5), and we deduce the
inequality (4.6). But more generally a different method is needed.

Very generally, probabilities for histories are given by an
expression of the form (4.1). If the projections contained in the
chain operators $C_{\au}$ are exact projections, and either
fine-grained projections onto discrete variables ({\it e.g.}, spins),
or coarse-grained projections onto continuous variables ({\it e.g.},
as in Eq.(3.2)), then
the variables $\au$ labeling the alternatives form a discrete set,
and so there are a discrete (although possibly infinite) set of
probabilities $p(\au)$. This means that they possess an upper bound,
$ p(\au) \le p_{max} \le 1 $, and a lower bound on the information
follows trivially:
$$
I \ = \ - \sum_{\au} p(\au) \ln p(\au) \ \ge \ \ln \left( {1 \over
p_{max} } \right)
\eqno(5.1)
$$
(Note that this is not true of the information of continuous
variables. There, the $p(\au)$'s are not probabilities, but
probability densities, and so need not be bounded from above.)
The upper bound $p_{max}$ may be computed by studying the spectrum
of the operator $C_{\au}^{\dag} C_{\au} $. In particular, the bound
(5.1) will be non-trivial, {\it i.e.}, $p_{max} < 1$, if at least
one pair of the time-dependent projections $P_{\a_k}^k(t_k)$ do not
commute [\cite{srinivas1}].

Now consider the case of two-time histories. As stated, everything
in Section IV from (4.1) to (4.5) also holds in the case of exact
projections. Suppose we obtain the spectrum of the operator
$\Omega$, Eq.(4.3), and we look for the largest eigenvalue, $\lambda_{max}$,
thus $\lambda_a \le \lambda_{max} $. It follows from (4.4)
that
$$
p(\a,\b)  \le \lambda_{max}
\sum_a \ \ \la a | \ U(\k,\x) \ \rho \ U^{\dag} (\k,\x) \ | a \ra
= \lambda_{max}
\eqno(5.2)
$$
and thus
$p_{max} = \lambda_{max}$. Position and momentum projections at the
same time, or position projections at different times do not
commute, thus the bound (5.1) will be non-trivial.

\subhead{\bf V(A). Position and Direct Momentum Samplings}

Consider first the case of a position followed by momentum sampling,
so we have (4.20), but with exact projections. We again deduce
(4.25), so we are interested in the spectrum of the operator
$\Omega$, given by (4.24). Now write
$$ \Omega | u
\ra = \lambda | u \ra
\eqno(5.3)
$$
Inserting the explicit form of
$\Omega$, and performing the $k$ integration, one obtains the
eigenvalue equation
$$
\theta(\xi) \theta(1- \xi) \int_0^1 d \xi'
\ { \sin\left(\pi U (\xi - \xi') \right) \over \pi
(\xi - \xi')} \ w(\xi') \ = \ \lambda w(\xi)
\eqno(5.4)
$$
where $U = \s_k \s_x / 2 \pi \hbar $, $x= \s_x (\xi-\half)$,  and
$w(\xi) = \la x | u \ra $. Apart from $\theta$-functions on the
left-hand side, this  equation is identical to an eigenvalue
equation written down by Partovi in his study of the analagous
question for the case of samplings of two ensembles, as in Section
III(B) [\cite{partovi1}]. It is not clear whether it can be solved
exactly, but it is straightforward to extract the relevant
information in regimes of interest. For $U <<1 $, the kernel on the
left-hand side is approximately equal to $ U $. The spectrum is
degenerate with $\lambda \approx U$, and $w(\xi)$ a constant on the
interval $[0,1]$ and zero elsewhere.
For $U>>1$, the kernel becomes a delta-function, $\delta(\xi -\xi')$.
The eigenvalue equation is then satisfied by any function with
support only in the interval $[0,1]$ (up to normalization), and
the spectrum is again degenerate with $\lambda \approx 1$.
The following bound on the information is thus obtained:
$$
I(K,X) \ \ge I_{min} \ \approx  \ \cases{0, &if $\s_k \s_x >> 2 \pi
\hbar$; \cr \ln \left( { 2 \pi \hbar \over \s_k \s_x } \right), &if
$\s_k \s_x << 2 \pi \hbar$. \cr}
\eqno(5.5)
$$

Like the continous case, (4.6), the result is intuitively appealing.
The lower bound is non-trivial for probes of phase space comparable
to or smaller than the fundamental cell. On  the other hand, there
is no restriction when the probe is much larger than the fundamental
cell, and the lower bound is essentially zero. (It is not $-\infty$,
as in the continuous case, because information is
non-negative for discrete distributions).

Note that the bounds (5.5) and (4.6) approximately coincide
for the case $ \s_k \s_x << 2 \pi \hbar $. This is to be expected
since as stated above, this is the condition that the discrete
and continous version of (4.5) coincide.

\subhead {\bf V(B). Position and Time-of-Flight Momentum Samplings}

In the case of time-of-flight momentum samplings, we study (4.26)
with exact position projections. We again have (4.33) and we thus
need to find the largest eigenvalue of the operator $\Omega$, in
this case given by (4.32). It is straightforward to show that the
eigenvalue equation is,
$$
\theta(\xi) \theta(1- \xi) \int_0^1 d \xi'
\ \exp \left( -i  \pi U (\xi -\xi') (\xi + \xi' +1 ) \right)
\ { \sin\left(\pi U (\xi - \xi') \right) \over \pi (\xi - \xi')} \ w(\xi')
\ = \ \lambda w(\xi)
\eqno(5.6)
$$
where the various quantities are all the same as in (5.4), recalling
that $\s_k = m \s_x / t $, as in Section IV(C). It is not difficult
to see that the presence of
the exponential factor in (5.6) in comparison to (5.4) actually
makes no difference to the leading order asymptotic solutions in the
regions $U>>1$ and $U<<1$. We thus once again obtain the result
(5.5).

As in Eq.(4.34), one can again use this result to
obtain a bound on the information of the joint probability of
position samplings. In this case it is,
$$
I(X_1,X_2) \ge I_{min} \approx \ln \left( { 2 \pi \hbar t \over
m \s_x^2 } \right)
\eqno(5.7)
$$
in the regime $ m \s_x^2 << 2 \pi \hbar t $. Similarly, we expect
to be able to derive a result of the form (4.42), for linear systems,
in the exact
projections case, although we do not describe this in detail.

\head {\bf VI. General Histories}

We have studied the uncertainty principle for histories
characterized by position and momentum projections at two moments of
time. We now go on to study the more general case of histories
characterized by position projections at an arbitrary number of
times [\cite{caves}].
On general grounds, and inspired by specific calculations
[\cite{dowker}], we expect the probability for a sequence of
position samplings to be in some sense
peaked about sets of solutions to the classical field equations,
with a weight depending on the initial state. The precise sense in
which this is true is discussed in another paper [\cite{jjh3}]
(see also Ref.[\cite{mensky1}]).
One expects the uncertainty principle to impose a limitation on the
degree of peaking. Here, we derive an information-theoretic
inequality expressing this limitation for histories characterized
by an arbitrary number of position samplings. This is a
generalization of the results (4.34), (4.42) and (5.7). We then
obtain the form of the uncertainty principle for histories
characterized by other types of projections.

As in Section V, if some of the projections in the chain operators
$C_{\au}$ do not commute, then the spectrum of the operator $C_{\au}
C_{\au}^{\dag}$ is strictly less than $1$, and likewise the
probabilities $p(\au)$. A lower bound on the information of
the form (5.1) is thus obtained.
Let us apply this rationale to strings of imprecise
position projections, with sampling functions of the form (3.2).
Our aim is to obtain a lower bound on the information
$$
I(X_1, \cdots X_n) = - \sum_{\a_1} \cdots \sum_{\a_n} \ p(\a_1 \cdots
\a_n) \ln p(\a_1 \cdots \a_n)
\eqno(6.1)
$$

The expression (1.3) for the probabilities may be written,
$$
p(\au ) = \int dx_0 dy_0
\ \la y_0 | C_{\au}^{\dag} C_{\au} | x_0 \ra
\ \rho(x_0, y_0)
\eqno(6.2)
$$
where
$$
\eqalignno{
\la y_0 | C_{\au}^{\dag} C_{\au} | x_0 \ra
=& \int
\ \prod_{k=1}^{n} dx_k dy_k \ \delta(x_n - y_n )
\ \U(x_k - \x_k) \U(y_k -\x_k)
\cr & \times
\prod_{k=1}^{n}
\ J(x_{k}, y_{k}, t_{k} | x_{k-1}, y_{k-1}, t_{k-1} )
&(6.3) \cr }
$$
Here, as in previous sections, $ \x_k = \s \a_k $. The samplings
functions $\U$ are given by Eq.(3.2). $J$ is the
density matrix propagator, which for unitary evolution is given by
$$
J(x^{\pp},y^{\pp},t^{\pp} | x',y',t') =
\la x^{\pp},t^{\pp} | x',t' \ra
\ \la y^{\pp},t^{\pp} | y',t' \ra^*
\eqno(6.4)
$$
We shall work in the limit that the time separation between each
projection is small. The propagators in (6.4) are then given by
(4.35), (4.36). This is exact for linear systems.
The case in which $J$ is a non-unitary reduced density matrix
propagator is also of interest in the context of decoherence models
(see Refs.[\cite{dowker,caldeira}], for example). However, such propagators
reduce to the unitary expression (6.4) in the short time limit, hence
our results are applicable to that case also.

For simplicity, we study first the free particle case, for which
one has
$$
\Delta(t^{\pp},t') =
\left( { m \over 2 \pi i \hbar (t^{\pp}-t') } \right)^{\half}
\eqno(6.5)
$$
and
$$
S(x^{\pp},t^{\pp}|x',t') = {  m (x^{\pp} -x')^2
\over 2 (t^{\pp}-t') }
\eqno(6.6)
$$
Also, let all of the projections have the same width $\s$, and let
the time separation between all slits be $t$ (except for $t_1$ and
$t_0$ -- see below).

We wish to estimate the largest eigenvalue of the operator
$C_{\au}^{\dag} C_{\au}$. The eigenvalue equation is,
$$
\int dx_0 \ \la y_0 | C_{\au}^{\dag} C_{\au} | x_0 \ra \ u(x_0)
= \lambda u(y_0)
\eqno(6.7)
$$
The expression (6.3) occurring in (6.7) has the form of a discrete
version of a sum over histories. It may be regarded as a sum
over pairs of paths, starting at $x_0$ and $y_0$, passing through
gates of width $\s$ at times $t_1 \cdots t_n$, and meeting in the
final gate at point $x_{n}$, which is integrated over the width $\s$.
We may
approximately evaluate (6.3), and hence solve the eigenvalue
equation, by looking for the paths which dominate the integral in
the regimes of interest.

We follow a heuristic argument previously used by Mensky
in a related context [\cite{mensky1}].
There are two competing effects that will
determine which paths dominate. On the one hand, if the slit widths
in the projections are very small, this will force the
paths to follow the set of alternatives $\x_k$ specified by the
projections. On the other hand, if the action of each path ({\it
i.e.} the sum of the phases of the propagators)
is very large, $S >> \hbar$, then by the
stationary phase approximation, we expect the dominant paths to be
those extremizing the action, {\it i.e.}, classical paths.

Consider first the case in which the slit widths are very small. In
this case the paths are forced to follow the sampling positions
$\x_k$. The action $S$ of each path is
of order $ m \s^2 / t $. We therefore take ``$\s$ small'' to mean
that $ S << \hbar $. This implies that the  exponential part of
the propagators in (6.3) is negligible, and only the prefactors
contribute. We may therefore approximately evaluate the integral
(6.3), with the result
$$
\eqalignno{
\la y_0 | C_{\au}^{\dag} C_{\au} | & x_0 \ra \  \approx
\ \s \ (\s^2)^{n-2} \  \left( {m \over 2 \pi \hbar t } \right)^{n-1}
\cr & \times
\ \int dx_1 dy_1 \U(x_1 -\x_1) \U(y_1 -\x_1)
\ J(x_1,y_1,t_1|x_0,y_0,t_0)
&(6.8) \cr }
$$
The origin of each part of this expression is as follows: the factor
$( m / 2 \pi \hbar t)^{n-1} $ comes from the $(n-1)$ propagators
$J$; the factor $(\s^2)^{n-2}$ comes from the integrations over $x$
and $y$ at times $t_2$ to $t_{n-1}$, noting that $J$ is
approximately constant, and recalling Eq.(3.4); the factor of $\s$
comes from the final integration over $x_n$.
The remaining integrations over $J$
in (6.8) arise due to the fact that the density
matrix in (6.2) is at the initial time $t_0$, and not at the time
$t_1$ at which the first projection is made. This is merely a
notational inconvenience -- the very last part of the chain
operators $C_{\au}$ is an evolution operator from $t_0$ to $t_1$.
It is readily removed by letting $t_1 \ria t_0$; thus $J$ becomes
a product of delta-functions and (6.8) becomes,
$$
\la y_0 | C_{\au}^{\dag} C_{\au} | x_0 \ra \ \approx
\ \s^{-1} \ \left( {m \s^2 \over 2 \pi \hbar t } \right)^{n-1}
\ \U( x_0 - \x_1) \U (y_0 - \x_1 )
\eqno(6.9)
$$
Inserting this in the eigenvalue equation (6.7), one thus finds that
the spectrum is degenerate, with
$$
\lambda \ \approx
\ \left( {m \s^2 \over 2 \pi \hbar t } \right)^{n-1}
\eqno(6.10)
$$
The eigenfunctions are functions constant in an interval of size
$\s$ and zero elsewhere.

Next, let the slit widths be very large. The action of each
section of path is then allowed to be large, and it is the
stationary phase effect that will dominate. The dominant
contribution to the sum over histories will therefore come from the
immediate vicinity of the classical paths. When the sampling
positions are chosen to line up according to the classical path,
it is as if the projections are not there, since most of the
integral comes from this regime anyway. It follows that
$$
\la y_0 | C_{\au}^{\dag} C_{\au} | x_0 \ra \ \approx \delta(x_0 -
y_0)
\eqno(6.11)
$$
and we thus find that
$\lambda_{max}  \approx 1 $.

Combining these two cases,
we thus obtain the following for the lower bound on the information
of a sequence of position samplings,
$$
I(X_1,X_2, \cdots X_n) \ge I_{min} \approx
\cases{0, &if $ m \s^2 >> \hbar t $;
\cr (n-1) \ln \left( { 2 \pi \hbar t \over m \s^2  } \right), &if
$m \s^2 << \hbar t $. \cr}
\eqno(6.12)
$$

This lower bound is what one might intuitively expect.
First of all, large $\s$ is
essentially the classical regime, in which we do not expect to
suffer limitations on our ability to describe a history;
hence there is no restriction on the information.
Secondly, the case of small $\s$ is essentially (4.34) generalized
to an arbitrary number of samplings. We might expect it because when
$\s$ is small, the projectors are almost fine-grained. They ``pinch
off'' the probability (6.2) -- it becomes approximately equal to a
product of probabilities for two-time histories of the type
discussed in Sections IV and V. Indeed, the bound in (6.12) is just a
sum of bounds of the type (4.34). We will see this in more detail below.

Generalizations of (6.12) may be obtained. The above analysis
is readily generalized to the case in which the slits widths $\s_j$
and the time separations $(t_{j+1}-t_j)$ are different, and the short
time propagator is given by the more general expression (4.35). It
is then straightforward to show that the lower bound in (6.12) is,
in the small $\s_j$ regime,
$$
I_{min} \approx - \sum_{j=1}^{n-1} \ln \left( \s_{j+1} \s_{j}
\bigl| \Delta(t_{j+1},t_{j}) \bigr|^2 \right)
\eqno(6.13)
$$
(and again $I_{min} \approx 0$ in the large $\s_j$ regime).
Eq.(6.13) is the leading order behaviour of $I_{min}$ for small
$\s_j$, and for small time separations. For linear systems it is
valid for arbitrary time separations.
How are we to understand this expression?

For the phase space samplings considered earlier, the significance
of the lower bounds (4.6), (5.5), is intuitively clear: the argument
of the logarithm is the ratio of the fundamental phase space volume
$2 \pi \hbar$ to  the sampling volume $\s_x \s_k$.

The lower bound (6.13) has a rather different form; yet an
analagous interpretation suggests itself. The propagator prefactor
$\bigl| \Delta(t_{j+1},t_{j}) \bigr|^2 $
has the dimension of $({\rm length})^{-2}$
and is commonly regarded as the ``density of paths''. Introduce the
quantity,
$$
V_{H} = \prod_{j=1}^{n-1} \ \bigr| \Delta(t_{j+1},t_{j}) \bigl|^{-2}
\eqno(6.14)
$$
for $n=2,3 \cdots$. For the case of position samplings
it has the dimension $({\rm length})^{2n-2}$. It might therefore reasonably
be regarded as the fundamental ``history space volume''.
Eq.(6.13) may then be written in the suggestive form,
$$
I_{min} \approx \ln \left( { V_H \over \s_1 \s_2^2 \cdots \s_{n-1}^2
\s_n} \right)
\eqno(6.15)
$$
Eq.(6.15) now has exactly the same structure as the
information-theoretic bounds (4.6), (5.5),
on the phase space samplings considered
earlier: the argument of the logarithm in (6.15) is the ratio of the
fundamental history space volume to the sampling volume.

It is natural to ask how the results of this section might be
further generalized to histories characterized by samplings of
variables other than position. It is actually not difficult to see
that the above results generalize to histories characterized by
samplings of any
continuous quantity, such as momentum, angular momentum, {\it
etc}. Let $P_{\bar \a} $ be an imprecise sampling of some continous
quantity $\a$:
$$
P_{{\bar \a}} = \int_{\s} d \a \ | \a \ra \la \a |
\eqno(6.16)
$$
The projections partition the variable $\a$ into bins of size $\s$
labeled by ${\bar \a}$. We may take the projections to be onto different
variables at each moment of time. In the limit of small widths, it
is not difficult to see that the analysis for the position samplings case
described above readily goes over to the case of arbitrary
continuous variables $\a_k$. Essentially what happens is that in
the small $\s_j$ limit, the matrix elements of the operator
$C_{\au}^{\dag}C_{\a}$ become products of propagators and slit
widths, in analogy with Eq.(6.8). More precisely,
$$
\eqalignno{
\la \a_0^{\prime} | C_{\au}^{\dag} C_{\au} | \a_0 \ra \ \approx
\ \int_{\s_1} d \a_1 & \int_{\s_1} d \a_1^{\prime}
\ \prod_{j=1}^{n-1} \ \s_{j+1} \Bigr| \la \a_{j+1}, t_{j+1} | \a_j,
t_j \ra \Bigl|^2 \s_j
\cr & \times
\ { 1 \over \s_1} \
\la \a_1, t_1 | \a_0, t_0 \ra
\ \la \a_1^{\prime}, t_1 | \a_0^{\prime}, t_0 \ra^*
&(6.17) \cr }
$$
We therefore again deduce the
lower bound on the uncertainty (6.15), for this much more general
class of histories. The factors $\Delta(t_{j+1},t_j)$ in (6.14) are now
identified with the short time limit of the progators
$ \la \a_{j+1},t_{j+1}|\a_j,t_j \ra $ (maximized over the
alternatives $\a_{j+1},\a_{j}$, in the event that the propagator
depends on them in the short time limit). We may thus write
$$
I(A_1, A_2, \cdots A_n) \ge I_{min} \approx
\ln \left( { V_H \over \s_1 \s_2^2 \cdots \s_{n-1}^2 \s_n } \right)
\eqno(6.18)
$$
in the small $\s_j$ regime.
Here, $A_1, A_2, \cdots A_n$ denotes a string of alternatives which
can be any continuous variables, and may be different variables at
different times.

Let us test this more general result with a simple case. Consider a
history characterized by a position projection at time $t_1$ and a
momentum projection at time $t_2$. Thus $\s_1 = \s_x$ and $\s_2 =
\s_k$. The short time propagator is
$$
\la p, t_2 | x, t_1  \ra \ \approx
\ {1 \over ( 2 \pi \hbar )^{\half} }
\exp \left( - { i p^2 (t_2- t_1) \over 2 m } - ipx \right)
\eqno(6.19)
$$

The history space volume is therefore  $V_H = \bigl| \Delta(t_2,t_1)
\bigr|^{-2} = 2 \pi \hbar $.
The history space volume element is not
just analogous to the factor of $2 \pi \hbar $ for phase space
samplings: it is equal to it in this case. Moreover, the
general result (6.18) coincides exactly with the expected result (5.5)
for phase space samplings.

Eq.(6.18) is the main result of this paper: a concise and very
general expression of the uncertainty principle, expressed in the
language of quantum-mechanical histories, not referring in any way to
phase space but reducing to the phase space form in the appropriate
circumstances.

The expression of the uncertainty principle (6.18) refers to a
fundamental history space volume $V_H$. It is obtained in (6.14)
from the
short time behaviour of the propagator, and is thus uniquely
determined given the unitary evolution operator, $e^{-\ih Ht}$. That
this operator should appear in the statement of the uncertainty
principle for histories should come as no surprise. Unlike phase
space statements, the description of a history depends on both the
projection operators at each moment of time {\it and} the unitary
evolution between them.

Of course, we have not defined the ``history space'' of which $V_H$
is the volume element. We shall not pursue this question here,
except to note that it appears to be related to the Cartesian
product space
$ s_1 \times s_2 \cdots s_n $, where $s_j$ is the
spectrum of the observable projected at time $t_j$.
This has been discussed by Omn\`es [\cite{omnes3}].
It is also perhaps
interesting to note that the existence and relevance of such a space
is indicated by the form of the uncertainty relation (6.18)

\head{\bf VII. Discussion}

In this paper, we addressed a simple question: How is the
uncertainty principle encoded in the probabilities for histories,
Eq.(1.3)? A simple but very general answer is offered:
it arises as the lower bound
on the Shannon information, Eq.(6.18).

We have stressed the generality of the lower bound (6.18) within the
framework of standard quantum mechanics (or at least, its modest
generalization to histories).
Yet the information-theoretic
approach employed here has a potentially greater degree of
generality. Information as a measure of uncertainty depends solely
on the probabilities for histories. This is in contrast to the usual
variance form of the uncertainty principle, (1.6), which depends on
the wave function of the system at a fixed moment of time. The
generality of the information-theoretic form suggests that it might
survive to broader forms of quantum mechanics, such as the
generalized quantum mechanics suggested by Hartle [\cite{hartle}],
which attempts to get away from the Hilbert space formulation. For
even if a formulation of quantum mechanics does not deal with wave
functions, it must deal with probabilities: information-theoretic
measures may therefore exist where Hilbert space-dependent
measures do not.

To be more precise, we
conjecture that the uncertainty principle will most generally arise
as a lower bound on the information, of the form (1.8), even in
generalized formulations of quantum mechanics in which a statement
in terms of variances is not available. A stronger conjecture is
that the general form of the lower bound (6.18) will also survive such
generalizations. These are, however, difficult issues to address in
the absence of a concrete generalization of quantum mechanics.
They will be taken up elsewhere.

\head{\bf Acknowledgements}

I am very grateful to numerous colleagues for useful conversations,
including  Arlen Anderson,
Carl Caves, Murray Gell-Mann, Jim Hartle, Salman Habib, Bei-Lok Hu,
Chris Isham,
Raymond Laflamme, Seth Lloyd, Warner Miller,
Roland Omnes, Juan Pablo Paz and Wojciech
Zurek. This work was supported by a
University Research Fellowship from the Royal Society.

\references

\def\pr{{\sl Phys. Rev.\ }}
\def\prl{{\sl Phys. Rev. Lett.\ }}
\def\prep{{\sl Phys. Rep.\ }}
\def\jmp{{\sl J. Math. Phys.\ }}
\def\rmp{{\sl Rev. Mod. Phys.\ }}
\def\cmp{{\sl Comm. Math. Phys.\ }}

\def\pl{{\sl Phys. Lett.\ }}
\def\annp{{\sl Ann. Phys. (N.Y.)\ }}

\def\ijtp{{\sl Int. J. Theor. Phys.\ }}

\def\jsp{{\sl J. Stat. Phys.\ }}

\refis{abe} S.Abe and N.Suzuki, \pr {\bf A41}, 4608 (1990).

\refis{ABL} Y.Aharanov, P.Bergmann and J.Lebowitz, \pr {B134},
1410 (1964).

\refis{anderson} A.Anderson, \pr {\bf D42}, 585 (1990).

\refis{ander} A.Anderson and J.J.Halliwell, ``An
Information-Theoretic Measure of Uncertainty due to Quantum and
Thermal Fluctuations'', Imperial College Preprint 92-93/25 (1993),
gr-qc 9304025.

\refis{beckner} W.Beckner, {\sl Ann.Math} {\bf 102}, 159 (1975).

\refis{bialy1} I.Bialynicki-Birula, \pl {\bf A103}, 253 (1984).

\refis{bialy2} I.Bialynicki-Birula and J.Mycielski, \cmp {\bf 44},
129 (1975).

\refis{busch} P.Busch and P.J.Lahti, {\sl J.Phys.} {\bf A20}, 899
(1987).

\refis{caldeira} A.O.Caldeira and A.J.Leggett, {\sl Physica} {\bf
121A}, 587 (1983).

\refis{caves} For an extensive discussion of quantum mechanics for
measurements distributed in time, see
C.Caves, \pr {\bf D33}, 1643 (1986).

\refis{cover} T.M.Cover and J.A.Thomas,
{{\it Elements of Information Theory}}
(Wiley, New York, 1991).

\refis{cover2} A.Dembo,T.M.Cover and J.A.Thomas, {\sl IEEE
Trans. Inform. Theory} {\bf 37}, 1501 (1991).

\refis{deutsch} D.Deutsch, {\prl} {\bf 50}, 631 (1983).

\refis{dowker} H.F.Dowker and J.J.Halliwell, \pr {\bf D46}, 1580
(1992).

\refis{gellmann}
M.Gell-Mann and J.B.Hartle, in {\it Complexity, Entropy and the
Physics of Information. SFI Studies in the Sciences of Complexity,
Vol. VIII}, edited by W.Zurek (Addison Wesley, Reading, MA, 1990);
M.Gell-Mann and J.B.Hartle, ``Classical Equations for Quantum
Systems'', UCSB preprint (to be published in \pr D, 1993).

\refis{hartle}
J.B.Hartle, in {\it Quantum Cosmology and Baby Universes:
Proceedings of the 1989 Jerusalem Winter School on Theoretical
Physics}, edited by S.Coleman, J.B.Hartle, T.Piran and S.Weinberg
(World Scientific, Singapore, 1991); and in ``Spacetime Quantum
Mechanics and the Quantum Mechanics of Spacetime'', preprint
UCSBTH92-91 (to appear in proceedings of the 1992 Les Houches Summer
School, {\it Gravitation et Quantifications}).

\refis{griffiths} R.Griffiths, \jsp {\bf 36}, 219 (1984).

\refis{grabowski2} M.Grabowski, \pl {\bf A124}, 19 (1987)

\refis{grabowski} M. Grabowski, {\sl Rep.Math.Phys.} {\bf 20}, 153 (1984).

\refis{everett} H.Everett in, {\it The Many-Worlds Interpretation of Quantum
Mechanics}, edited by B.S.DeWitt and N.Graham (Princeton University
Press, Princeton, NJ, 1973).

\refis{jjh} J.J.Halliwell, \pr {\bf D46}, 1610 (1992).

\refis{jjh3} J.J.Halliwell, ``Quantum-Mechanical Histories and the
Uncertainty Principle. II. The Peaking about Classical Histories'',
Imperial College Preprint (1993).

\refis{jjh4} J.J.Halliwell, in preparation.

\refis{hirsch} I.I.Hirschmann, {\sl Amer.J.Math.} {\bf 79}, 152
(1957).

\refis{klauder} J.R.Klauder and E.C.G.Sudarshan, {\it Fundamentals
of Quantum Optics} (Benjamin, New York, NY, 1968); J.R.Klauder and
B.S.Skagerstam, {\it Coherent States} ( World Scientific, Singapore,
1985).

\refis{lieb} E.H.Lieb, \cmp {\bf 62}, 35 (1978).

\refis{maassen} H.Maassen and J.B.M.Uffink, \prl {\bf 60}, 1103
(1988).

\refis{mamojka} B.Mamojka, \ijtp {\bf 11}, 73 (1974).

\refis{mensky1} M.Mensky, \pl {\bf A155}, 229 (1991).

\refis{omnes} R.Omn\`es, \rmp {\bf 64}, 339 (1992), and references
therein.

\refis{omnes2} See for example,  R.Omn\'es, {\sl J.Stat.Phys.} {\bf
57}, 357 (1989), for a review of microlocal analysis as it affects
expressions of the form (1.3).
For earlier work see, L.H\"ormander,
{\it The Analysis of Linear Partial Differential Equations}
(Springer, Berlin, 1985).

\refis{omnes3} R.Omn\`es, \annp {\bf 201}, 354 (1990).

\refis{partovi1} M.H.Partovi, \prl {\bf 50}, 1883 (1983).

\refis{partovi2} R.Blankenbecler and M.H.Partovi, \prl {\bf 54}, 373
(1985).

\refis{raja} A.K.Rajagopal and S.Teitler, \pl {\bf A115}, 313
(1986).

\refis{sanchez} J.S\'anchez, \pl {\bf A173}, 270 (1993).

\refis{schroeck} F.E.Schroeck, \jmp {\bf 30}, 2078 (1989).

\refis{shannon} C.E.Shannon and  W.W.Weaver,
{{\it The Mathematical Theory of Communication}}
(University of Illinois Press, Urbana, IL, 1949),

\refis{srinivas1} Lower bounds on the information obtained in this
way for arbitrary pairs of non-commuting projections have
previously been discussed by M.D.Srinivas, {\sl Pramana} {\bf 24}, 673
(1985), but the explicit form of the bounds was not computed.


\refis{wehrl} A.Wehrl, {\sl Rep.Math.Phys.} {\bf 16}, 353 (1979).

\refis{wigner} See, for example, E.P.Wigner, in {\it Quantum Theory
and Measurement}, edited by J.A.Wheeler and W.H.Zurek (Princeton
University Press, Princeton, NJ, 1983).

\refis{wignerfn} N.Balazs and B.K.Jennings, \prep {\bf 104}, 347 (1984),
M.Hillery, R.F.O'Connell, M.O.Scully and E.P.Wigner, \prep {\bf
106}, 121 (1984).

\endreferences

\end